\documentclass[pre,showpacs,showkeys,twocolumn,floatfix,amsmath,amsfonts]{revtex4}

\usepackage{graphicx}
\usepackage{epstopdf}
\usepackage{epsfig}
\usepackage{textcomp}
\usepackage[utf8]{inputenc}
\usepackage[english]{babel}
\usepackage[usenames, dvipsnames]{color}
\usepackage{color}
\usepackage{soul}

\usepackage[colorlinks]{hyperref}

\hypersetup{breaklinks=true,colorlinks=true,citecolor=blue,linkcolor=blue,urlcolor=blue}

\begin{document}

\title{Equilibration of sinusoidal modulation of temperature in linear and nonlinear chains}% Force line breaks with \\

\author{Elena~A.~Korznikova$^{1,2}$}
\email{elena.a.korznikova@gmail.com}
\author{Vitaly~A.~Kuzkin$^{3,4}$}
\email{kuzkinva@gmail.com}
\author{Anton~M.~Krivtsov$^{3,4}$}
\email{akrivtsov@bk.ru}
\author{Daxing~Xiong$^5$}
\email{phyxiongdx@fzu.edu.cn}
\author{Vakhid~A.~Gani$^{6,7}$}
\email{vagani@mephi.ru}
\author{Aleksey~A.~Kudreyko$^{8}$}
\email{alexkudreyko@mail.ru}
\author{Sergey~V.~Dmitriev$^{1,9}$}
\email{dmitriev.sergey.v@gmail.com}

\affiliation{
$^1$Institute of Molecule and Crystal Physics, Ufa Federal Research Centre of the Russian Academy of Sciences, Ufa 450054, Russia\\
$^2$Ufa State Aviation Technical University, Ufa 450008, Russia\\
$^3$Peter the Great Saint Petersburg Polytechnical University, Saint Petersburg 195251, Russia\\
$^4$Institute for Problems in Mechanical Engineering, RAS, Saint Petersburg 199178, Russia\\
%$^5$Department of Physics, Fuzhou University, Fuzhou 350108, Fujian, China\\
$^5$MinJiang Collaborative Center for Theoretical Physics, Department of Physics and Electronic Information Engineering, Minjiang University, Fuzhou, Fujian 350108, China\\
$^6$National Research Nuclear University MEPhI (Moscow Engineering Physics Institute), Moscow 115409, Russia\\
$^7$Institute for Theoretical and Experimental Physics of National Research Centre ``Kurchatov Institute'', Moscow 117218, Russia\\
$^8$Department of Medical Physics and Informatics, Bashkir State Medical University, Ufa 450008, Russia\\
$^9$Institute of Mathematics with Computing Centre, Ufa Federal Research Centre of RAS, Ufa 450008, Russia
}


\begin{abstract}
The equilibration of sinusoidally modulated distribution of the kinetic temperature is analyzed in the $\beta$-Fermi-Pasta-Ulam-Tsingou chain with different degrees of nonlinearity and for different wavelengths of temperature modulation. Two different types of initial conditions are used to show that either one gives the same result as the number of realizations increases and that the initial conditions that are closer to the state of thermal equilibrium give faster convergence. The kinetics of temperature equilibration is monitored and compared to the analytical solution available for the linear chain in the continuum limit. The transition from ballistic to diffusive thermal conductivity with an increase in the degree of anharmonicity is shown. In the ballistic case, the energy equilibration has an oscillatory character with an amplitude decreasing in time, and in the diffusive case, it is monotonous in time. For smaller wavelength of temperature modulation, the oscillatory character of temperature equilibration remains for a larger degree of anharmonicity. For a given wavelength of temperature modulation, there is such a value of the anharmonicity parameter at which the temperature equilibration occurs most rapidly. \end{abstract}

%Use showkeys class option if keyword
\maketitle
	
%\tableofcontents

\section{Introduction}
\label{Introduction}

The miniaturization of electronic and microelectromechanical systems has led to the need to control heat fluxes at the micro- and nanolevels, where the laws established for macrosystems turned out to be inaccurate; for example, the Fourier law of thermal conductivity is not valid~\cite{Experiment01,Experiment02,Experiment03,Experiment04,Experiment05,Experiment06}. On the other hand, the possibility of creating nanostructured metamaterials with desired properties endows researchers and technologists with new levers of heat flow control. Heat in materials is transferred mainly by phonons, and {\em phononics} is a rapidly growing field of knowledge at the intersection of physics, materials science, and nanotechnology, studying phonon energy transport and its applications~\cite{Maldovan209,Li1045}. Major achievements in this field are the development of thermal transistors~\cite{Li143501,Joulain200601}, thermal diodes~\cite{Li184301,Liu236,Savin245415,Hu1602726,Duan3133}, and thermal logic devices~\cite{Wang177208,Fornieri258,Murad193109}. Such developments require a better theoretical and experimental understanding of anomalous thermal transport in miniature low-dimensional systems.

The Fourier law of thermal conductivity states that for macroscopic bodies, the heat flux is proportional to the temperature gradient with a proportionality coefficient $\kappa$, referred to as thermal conductivity. In a number of theoretical works~\cite{Lepri377,Dhar457,Chang305,Evazzade163,Kuzkin,Coolig12,Wang013179,You012125,Xiong012130,Xiong022116,Bhattacharyya2020,Sato012111,Wang012126}, it was shown that in one-dimensional structures, Fourier's law does not work, in the sense that $\kappa$ depends not only on the material, but also on the dimensions of the conductor, in particular, on its length $L$ according to the power law $\kappa \sim L^\alpha$, where the exponent is in the range $0\le \alpha\le 1$. The case $\alpha = 1$ corresponds to ballistic thermal conductivity, and for $\alpha = 0$ we have normal, diffusive thermal conductivity obeying the Fourier law. If $0<\alpha<1$, we have anomalous thermal conductivity. Note that defect-free linear systems of any complexity always demonstrate ballistic heat propagation~\cite{Dugar022131,Lin01,Lin02,Lin03,Lin04,Lin05,Krivtsov407,Lin06,Lin07}. One might expect that in systems with weak anharmonicity, the linear theory can be very helpful~\cite{BallistResonance}. %~\cite{Evazzade2017,Hadipour2020,Korznikova2020}.

It was shown that the ballistic propagation of heat in linear discrete systems can be described with high accuracy by continuum equations~\cite{Krivtsov407,BallistResonance}. In particular, equilibration of sinusoidal modulation of temperature was described analytically~\cite{Krivtsov407}, and it was shown that temperature oscillations in the $\alpha$-Fermi-Pasta-Ulam-Tsingou ($\alpha$-FPUT) chain lead to the excitation of mechanical vibrations. This physical phenomenon is referred to as the ballistic resonance~\cite{BallistResonance}. 

For the description of anomalous thermal conductivity, the nonlocal fractional-type diffusion equation~\cite{Dhar457} and nonlinear fluctuating hydrodynamics~\cite{Spohn1,Spohn2} have been developed.

The equilibration of sinusoidal modulation of temperature was studied earlier in anharmonic chains~\cite{Gendelman020103,Zhao042136}. The authors found that the equilibration of the short-wavelength modulation of temperature occurs through oscillations, while the long-wavelength modulation is equilibrated monotonically. Here we give explanations for those observations, demonstrating that the oscillatory regime of temperature equilibration can be described by the linear theory and that the oscillations of the long-wavelength modulations are suppressed by a weaker anharmonicity as compared to short-wavelength modulations.

Here the $\beta$-FPUT chain with symmetric anharmonicity is considered~\cite{FPU} because it is free of thermal expansion and the mechanical and thermal oscillations are not coupled~\cite{BallistResonance}.

The main goal of this study is to analyze the transition from ballistic~\cite{Krivtsov407,BallistResonance} to diffusive~\cite{Dhar457,Gendelman020103,Zhao042136} energy equilibration as the degree of anharmonicity increases.

We describe the $\beta$-FPUT model and simulation setup in Sec.~\ref{SimulationSetup}, numerical results are presented in Sec.~\ref{Results}, and conclusions are drawn in Sec.~\ref{Conclusion}.

\section{The model and simulation setup}
\label{SimulationSetup}

\subsection{The Fermi-Pasta-Ulam-Tsingou chain}
\label{Sec_FPU}

We consider the $\beta$-FPUT chain of particles \cite{FPU} having mass $m$, whose dynamics is defined by the Hamiltonian
\begin{eqnarray}\label{Hamiltonian}
H = K + P = \sum_n K_n^{} + \sum_n P_n^{},
\end{eqnarray}
which is the sum of the kinetic ($K$) and potential ($P$) energies of the chain with the kinetic, potential, and total energies of individual particles being equal to
\begin{eqnarray}\label{Kn}
K_n^{} &=& \frac{m\dot{u}_n^2}{2}, \\ \label{Pn}
P_n^{} &=& \frac{k}{4}\left(u_n^{}-u_{n-1}^{}\right)^2 + \frac{\beta}{8}\left(u_n^{}-u_{n-1}^{}\right)^4 \nonumber\\ \label{Hn}
&+& \frac{k}{4}\left(u_{n+1}^{}-u_n^{}\right)^2 + \frac{\beta}{8}\left(u_{n+1}^{}-u_n^{}\right)^4,\\
H_n^{} &=& K_n^{} + P_n^{},
\end{eqnarray}
respectively. Here $u_n^{}(t)$ is the displacement of the $n$th particle from its equilibrium position, which is an unknown function of time $t$, and $\dot{u}_n^{}\equiv du_n^{}/dt$ is its velocity. The particles are coupled to their nearest neighbors by the potential which includes the quadratic term with the harmonic force constant $k$ and the quartic term with the anharmonic force constant $\beta$. 

The equations of motion that stem from Eqs.~\eqref{Hamiltonian}--\eqref{Pn} are
\begin{eqnarray}\label{EMo}
m\ddot{u}_n^{} = k\left(u_{n-1}^{}-2u_n^{}+u_{n+1}^{}\right) \nonumber\\
- \beta\left(u_n^{}-u_{n-1}^{}\right)^3 + \beta\left(u_{n+1}^{}-u_n^{}\right)^3.
\end{eqnarray}

Without loss of generality, we set $m$=1, $k=1$, and take different values for $\beta$. The lattice spacing is $h=1$. The chain of $N$ particles with the periodic boundary conditions ($u_n^{}=u_{n+N}^{}$) is considered.

In the case of small amplitude vibrations, one can neglect the nonlinear term by setting $\beta=0$ and find the solutions of the linearized equation \eqref{EMo} in the form of normal modes $u_n^{} \sim \exp [i (2\pi q n/N -\omega_{\rm q}^{} t)]$ with the wave number $q=0,1,...,N/2$ and frequency $\omega_{\rm q}^{}$. These modes obey the following dispersion relation:
\begin{equation}\label{Dispersion}
\omega_{\rm q}^{} = 2\:\sqrt{\frac{k}{m}}\sin\frac{\pi q}{N}.
\end{equation}
The considered chain supports the small-amplitude running waves (phonons) with frequencies within the band from $\omega_{\min}^{}=0$ for $q=0$ to $\omega_{\max}^{}=2\sqrt{k/m}$ for $q=N/2$.

As a measure of temperature, the averaged kinetic energy per atom,
\begin{equation}\label{Ken}
T=\bar{K}=\frac{1}{N}\sum_n\frac{m\langle\dot{u}_n^2\rangle}{2},
\end{equation}
is used, where $\langle \cdot \rangle$ denotes the mathematical expectation.

\subsection{Two types of initial conditions}
\label{Sec_IC}

Since we deal with temperature and take a relatively small number of particles, the physical picture emerges as a result of averaging over many realizations. 

We aim to set the initial conditions that create, after averaging over many realizations, the initial distribution of the total energy over the particles of the form
\begin{equation}\label{H_distr}
\langle H_n^{}\rangle = H_{\rm b}^{} + \epsilon\Big[1+\sin \Big(\frac{2\pi n}{N}\Big)\Big],
\end{equation}
where $H_{\rm b}^{}$ is the background level of the total energy and the term with a multiplier $\epsilon \ll H_{\rm b}^{}$ adds the desired sinusoidal modulation. 

The initial total energy distribution \eqref{H_distr} can be achieved in many different ways. We will compare the results for two types of initial conditions. The small sinusoidal addition, in both cases, will be taken in the form of kinetic energy as described below. On the other hand, the background energy, which constitutes the main part of the total energy of the system, will be introduced either as kinetic energy or it will be nearly equally shared between the kinetic and potential forms. 

{\em Initial conditions of the first type.} We assign random velocities to the particles with zero initial displacements so that the initial total energy includes only kinetic energy, while the potential energy is zero. To do so, in each realization, we set the initial velocities of the particles such that
\begin{equation}\label{v_n}
\dot{u}_n^{} = \rho_n^{} \sqrt{\frac{2}{m}} \sqrt{H_{\rm b}^{} + \epsilon\Big[1 + \sin\Big(\frac{2\pi n}{N}\Big)\Big]},
\end{equation}
where $\rho_n^{}$ is a random variable with the standard normal distribution having zero expectation and unit variance. For generation of random variable $\rho_n^{}$, the standard normal distribution given by the probability density function $P(x)=\exp(-x^2/2)/\sqrt{2\pi}$ is used. Then the desired spatial distribution of the total energy at $t=0$ is achieved,
\begin{equation}\label{v_n2}
\langle H_n^{}\rangle = \langle K_n^{}\rangle = \frac{m}{2}\langle \dot{u}_n^2\rangle =  H_{\rm b}^{} + \epsilon \Big[1 + \sin\Big(\frac{2\pi n}{N}\Big)\Big].
\end{equation}
%

%{\em The second type of initial conditions.} Here the background energy level $H_b$ will be achieved by summing up all $N/2$ standing waves 
%
%\begin{equation}\label{u_n}
%u_n =  b_{N/2}\cos(\pi n)+
%\sum_{q=1}^{N/2-1}b_q\cos\Big( %\frac{2q\pi n}{N}+\delta_q \Big),
%\end{equation}
%
%with random phase shifts $\delta_q$ uniformly distributed in the domain $(0,2\pi)$, and with the amplitudes $b_q$ chosen such that each harmonic has same total energy equal to $H_b/(N/2)$. As such, the background energy will be in the form of potential energy with zero kinetic energy. The desired energy modulation will be introduces by setting initial velocities of particles according to
%
%\begin{equation}\label{v_nsec}
%\dot{u}_n = \rho_n\sqrt{\frac{2}{m}} 5\sqrt{ \epsilon %\Big[1+\sin\Big(\frac{2\pi %n}{N}\Big)\Big]},
%\end{equation}
%
%This will produce total energy mainly in the form of the potential energy with a small sinusoidally distributed kinetic energy,
%
%\begin{equation}\label{v_n2sec}
%\langle H_n\rangle=\langle %K_n\rangle+\langle P_n\rangle= %\epsilon \Big[1+\sin\Big(\frac{2\pi %n}{N}\Big)\Big]+H_b.
%\end{equation}
%

{\em Initial conditions of the second type.} In this case, the background total energy $H_{\rm b}^{}$ is obtained by summing up all $N/2$ running harmonics \cite{Samarskii},
\begin{equation}\label{u_n}
u_n^{} =  b_{N/2}^{}\cos(\pi n) %\nonumber \\
+ \sum_{q=1}^{N/2-1}b_{\rm q}^{}\cos\Big( \frac{2q\pi n}{N}\pm \omega_{\rm q}^{} t+\delta_{\rm q}^{} \Big),
\end{equation}
with $\omega_{\rm q}^{}$ given by Eq.~\eqref{Dispersion}, with random phase shifts $\delta_{\rm q}^{}$ uniformly distributed in the domain $(0,2\pi)$, and with the amplitudes $b_{\rm q}^{}$ chosen such that each harmonic has same total energy equal to $H_{\rm b}^{}/(N/2)$. A plus or minus sign in front of the $\omega_{\rm q}^{} t$ term is taken with equal probability in order to have equal contribution to the energy from the waves running to the right and to the left.

For running waves in the linear lattice ($\beta=0$), the kinetic and potential energies are exactly equal, while for $\beta>0$, there appears a deviation from equality due to the effect of nonlinearity.

The sinusoidal modulation of the energy distribution along the chain is achieved for each realization by increasing the particle kinetic energies by
\begin{equation}\label{Add_Kn}
\Delta K_n^{} = \epsilon\Big[1+\sin \Big(\frac{2\pi n}{N}\Big)\Big],
\end{equation}
so that each particle having velocity $\dot{u}_n^{}$ gets the velocity increment,
\begin{equation}\label{Add_v}
\Delta \dot{u}_n^{} = \pm\sqrt{\dot{u}_n^2 + \frac{2}{m}\Delta K_n^{}}-\dot{u}_n^{},
\end{equation}
where the upper (lower) sign is for positive (negative) $\dot{u}_n^{}$.

{\em Summary}. In the initial conditions of the first type, the total energy of the chain at $t=0$ is in the form of kinetic energy with the potential energy being exactly zero. In the initial conditions of the second type, the background energy at $t=0$ is almost equally shared between the kinetic and potential forms (exactly equally in the linear case) and a small amount of the kinetic energy is added to achieve the desired sinusoidal modulation of the total energy distribution. 

\subsection{Analytical solution for the linear chain}
\label{Sec_Linear}

Heat transfer in the linear chain can be described with a high accuracy by the continuum equation~\cite{Krivtsov407}
\begin{equation}\label{ContEq}
\ddot{T} + \frac{1}{t}\dot{T} = c^2T^{\prime\prime},
\end{equation}
where $T(x,t)$ is the temperature field and 
\begin{equation}\label{SoundVel}
c = h\:\sqrt{\frac{k}{m}},
\end{equation}
is the sound velocity. 

For the initial conditions
\begin{equation}\label{ContInit}
T = T_{\rm b}^{} + \Delta T\sin(\lambda x), \quad \dot{T} = 0,
\end{equation}
where
\begin{equation}\label{lambda}
\lambda = \frac{2\pi}{L},
\end{equation}
and $L$ is the modulation wavelength, the solution to Eq.~\eqref{ContEq} reads~\cite{Krivtsov407}
\begin{equation}\label{ContSolut}
T = T_{\rm b}^{} + A(t)\sin(\lambda x), \quad A(t) = \Delta T J_0^{}(\omega t),
\end{equation}
where 
\begin{equation}\label{OmegaSolut}
\omega = \lambda c,
\end{equation}
and $J_0^{}$ is the Bessel function of the first kind.

Initially, the left half of the chain, $0\le x\le \pi/\lambda$, has an averaged temperature greater than the right half, $\pi/\lambda< x\le2\pi/\lambda$. Temperature equilibration in the system can be monitored by comparing average temperatures of the left and right halves,
\begin{equation}\label{TLTR}
T_{\rm L}^{} = \frac{\lambda}{\pi}\int_{0}^{\pi/\lambda} T dx, \quad T_{\rm R}^{} = \frac{\lambda}{\pi}\int_{\pi/\lambda}^{2\pi/\lambda} T dx.
\end{equation}
Substituting Eq.~\eqref{ContSolut} into Eq.~\eqref{TLTR}, we obtain the difference between the averaged temperatures,
\begin{equation}\label{TLmTR}
\delta T = T_{\rm L}^{} - T_{\rm R}^{} = \frac{4}{\pi} \Delta T J_0^{}(\omega t).
\end{equation}

A comparison of Eq.~\eqref{ContInit} with Eq.~\eqref{H_distr}, taking into account that kinetic energy is equal to one-half of total energy, suggests that
\begin{equation}\label{Tb}
T_{\rm b}^{} = \frac{1}{2} (H_{\rm b}^{}+\epsilon), \quad \Delta T = \frac{\epsilon}{2}, \quad \lambda = \frac{2\pi}{Nh}.
\end{equation}
In view of Eqs.~\eqref{SoundVel}, \eqref{OmegaSolut}, and \eqref{Tb}, we rewrite Eq.~\eqref{TLmTR} in terms of the parameters of our computational model,
\begin{equation}\label{TLmTRfin}
\delta T=T_{\rm L}^{} - T_{\rm R}^{} = \frac{2 \epsilon}{\pi} J_0^{}\left(\frac{2\pi}{N} \sqrt{\frac{k}{m}}\: t\right).
\end{equation}

\subsection{Simulation protocol}
\label{Sec_SP}

There are two ways to address the effect of anharmonicity on the dynamics of the chain. The first approach is to fix the anharmonicity parameter $\beta$ and study systems with different levels of energies (different temperatures). The second approach is to fix the energy of the system and to change $\beta$. In the present study, the second method is used.

The equations of motion given by Eq.~\eqref{EMo} are integrated numerically using the symplectic, sixth-order St\"ormer method with the time step $\tau=10^{-3}\sqrt{m/k}$. The accuracy of integration is controlled by monitoring the total energy of the system which, in our simulations, is conserved with the relative error not exceeding $10^{-7}$ within the whole numerical run.

In the simulations, we take $H_{\rm b}^{}=1.0$ and $\epsilon=0.1$ and consider different values of the nonlinearity parameter $\beta$ including the case of the linear chain with $\beta=0$. The initial conditions are taken in one of two forms, as described in Sec.~\ref{Sec_IC}.

As mentioned above, initially the left half of the chain has an averaged temperature $T_{\rm L}^{}$ greater than the right half, $T_{\rm R}^{}$. In order to study the kinetics of temperature equilibration, the difference between these temperatures,
\begin{equation}\label{TLTRdiscrete}
\delta T = T_{\rm L}^{} - T_{\rm R}^{} = \frac{2}{N}\sum_{n=0}^{N/2-1}\langle K_n^{}\rangle - \frac{2}{N}\sum_{n=N/2}^{N-1}\langle K_n^{}\rangle ,
\end{equation}
is calculated as a function of time and compared to the prediction of the linear theory, given by Eq.~\eqref{TLmTRfin}. Note that Eq.~\eqref{TLTRdiscrete} is the discrete version of Eq.~\eqref{TLTR}.

We also carry out simulations for chains of different lengths $N$. We emphasize that in most calculations, the temperature modulation wavelength is equal to the chain length $N$ and, only at the end of Sec.~\ref{Anharmonic}, we consider the case of a chain including several temperature modulation wavelengths.

The period of the Bessel function $J_0^{}(\xi)$ gradually shortens, rapidly approaching the value of $2\pi$. Suppose we want to analyze the dynamics of the chain for about $I$ periods of the Bessel function, in other words, for $\xi\le 2\pi I$. Then, in view of Eq.~\eqref{TLmTRfin}, the simulation time is about
\begin{equation}\label{tsimul}
t_{\rm s}^{} = IN\sqrt{\frac{m}{k}}.
\end{equation}
Recall that in our simulations, we set $m=1$ and $k=1$.

\section{Simulation results}
\label{Results}

\subsection{Characteristics of the initial conditions}
\label{StatIniCond}

The initial conditions of the first type, given by Eq.~\eqref{v_n}, and of the second type, given by Eqs.~\eqref{u_n} and \eqref{Add_v}, are stochastic, and it is instructive to estimate their statistical characteristics for chains of different lengths $N$.

\begin{figure}
\includegraphics[width=6.0cm]{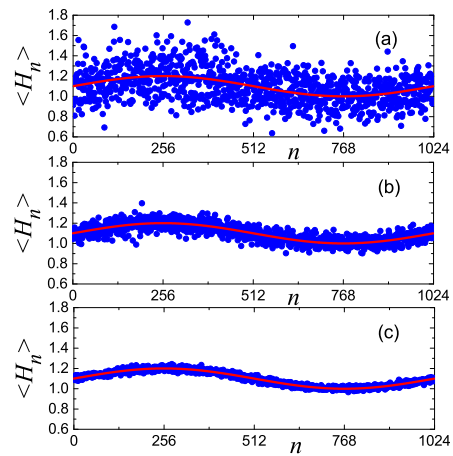}
\caption{Values of particle energies $\langle H_n^{}\rangle$ in the chain of $N=1024$ particles for the initial conditions of the first type averaged over (a) $10^2$, (b) $10^3$, and (c) $10^4$ realizations. The red line shows the desired energy distribution \eqref{H_distr} for $H_{\rm b}^{}=1$ and $\epsilon=0.1$. Here the chain is linear, $\beta=0$.}
\label{fig:fig1}
\end{figure}
\begin{figure}
\includegraphics[width=6.0cm]{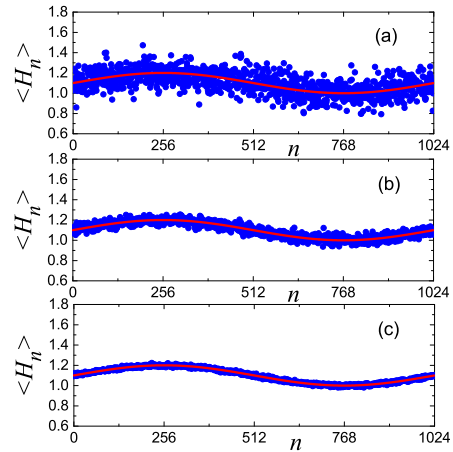}
\caption{The same as in Fig.~\ref{fig:fig1}, but for the initial conditions of the second type.}
\label{fig:fig2}
\end{figure}
\begin{figure}
\includegraphics[width=8.0cm]{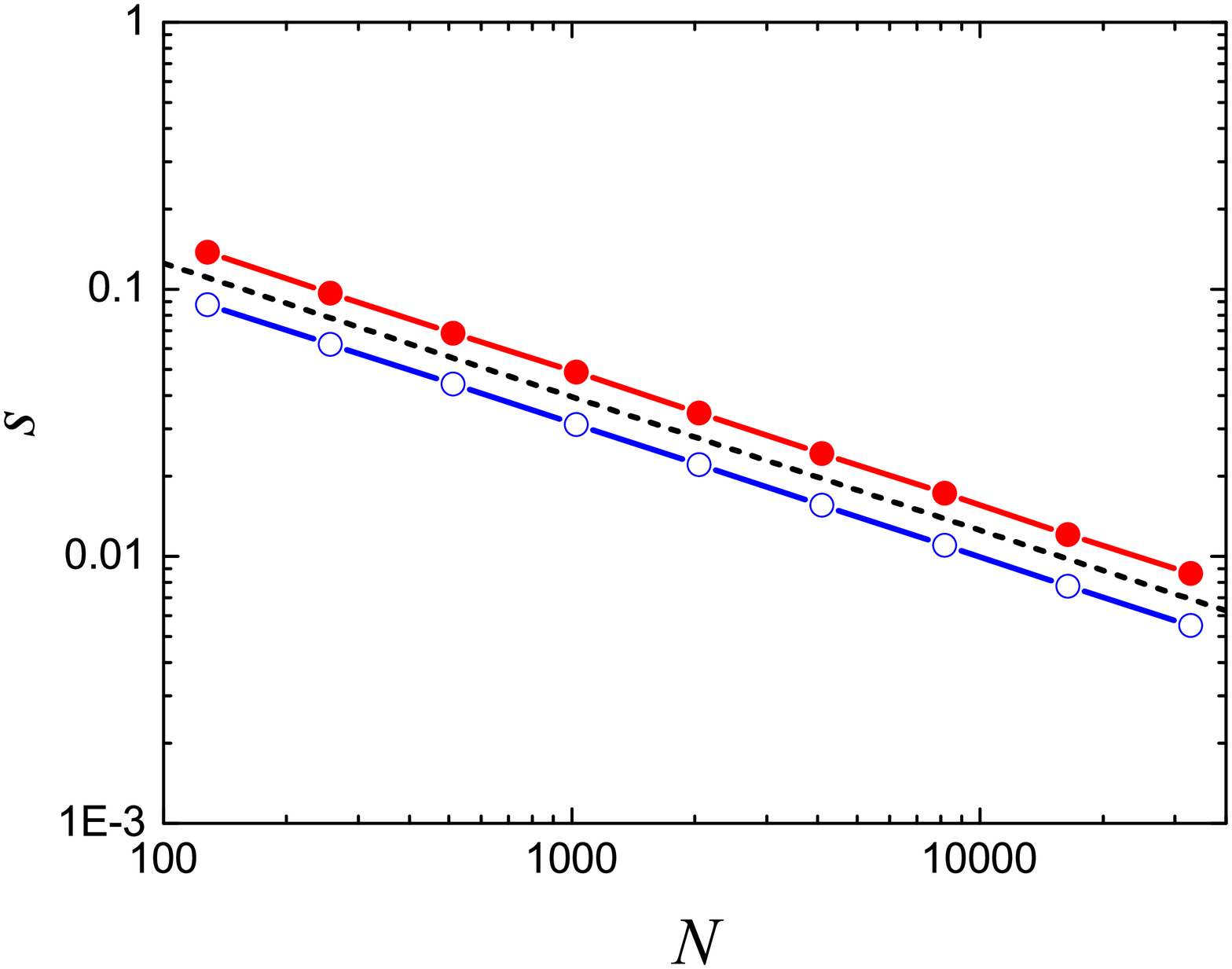}
\caption{Standard deviation of the difference between the temperatures of the left and right halves of the chain, $\delta T$, at $t=0$ calculated for $10^4$ realizations of the initial conditions of the first type (dots) and the second type (circles) for the chains of different length $N$. The dashed line shows the slope of $-1/2$. Results for $\beta=0$ (linear case).}
\label{fig:s}
\end{figure}

First, in Figs.~\ref{fig:fig1} and \ref{fig:fig2}, we show the initial distribution of total energy in the linear chain ($\beta=0$) of $N=1024$ particles for the initial conditions of the first and second types, respectively, for $H_{\rm b}^{}=1$ and $\epsilon=0.1$. The energies of the particles are averaged over (a) $10^2$, (b) $10^3$, and (c) $10^4$ realizations. The red line in each panel shows the desired distribution of energy, given by Eq.~\eqref{H_distr}. 

A comparison of the results presented in Figs.~\ref{fig:fig1} and \ref{fig:fig2} shows that the initial conditions of the second type are less stochastic than the initial conditions of the first type. This is understandable because in the initial conditions of the second type, the background energy $H_{\rm b}^{}$ is obtained by summation of phonon modes followed by correction of the particles' velocities to achieve the sinusoidal distribution of energy, while in the first type of the initial conditions, all particles have random initial velocities. 

In this work, energy equilibration in the chain will be monitored by observing the time evolution of an integral characteristic, namely, the difference between the temperatures of the left and right halves of the chain, $\delta T=T_{\rm L}^{}-T_{\rm R}^{}$. Let us analyze the statistical characteristics of $\delta T$ at $t=0$. According to Eq.~\eqref{TLmTRfin}, at $t=0$ one should have the mean value of $\delta T=2\epsilon/\pi$. We generate sets of $10^4$ initial conditions for different $N$ and calculate the standard deviation $s$ of $\delta T$ for $\beta=0$ (linear case). In Fig.~\ref{fig:s}, the standard deviation of $\delta T$ is presented as a function of the number of particles in the chain $N$ for the initial conditions of the first type (dots) and the second type (circles). The log-log plot shows that $s=a/\sqrt{N}$ (the offset dashed line has the slope $-1/2$) and $a=1.56$ for the initial conditions of the first type, while $a=0.989$ for the initial conditions of the second type. The results of Fig.~\ref{fig:s} confirm that the initial conditions of the second type are less stochastic because, for them, $s$ is smaller than for the initial conditions of the first type. 

\begin{figure}[t!]
\includegraphics[width=8.0cm]{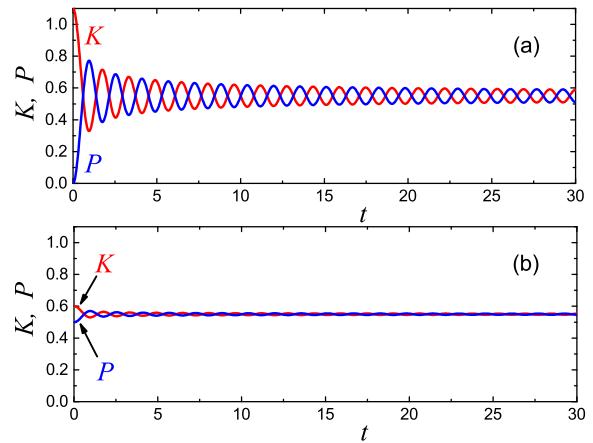}
\caption{Kinetic ($K$) and potential ($P$) energies of the chain as functions of time. Initial conditions of the (a) first and (b) second type are used. The presented result is the averaged one over 100 realizations. The linear chain ($\beta=0$) includes $N=2^{15}=32768$ particles.}
\label{fig:fig4}
\end{figure}

Recall that when the initial conditions of the first type are used, the potential energy of the chain at $t=0$ is zero and the total energy is equal to the kinetic energy. According to the exact result obtained for linear chains, the energy exchange between the kinetic and potential parts follows the Bessel function~\cite{Lin06}. In Fig.~\ref{fig:fig4}, we plot the time evolution of the kinetic and potential energies in the linear chain ($\beta=0$) of $N=2^{15}=32768$ particles for the initial conditions of the first [Fig.~\ref{fig:fig4}(a)] and second [Fig.~\ref{fig:fig4}(b)] types. The result is the average over 100 realizations. In the case of the first (second) initial conditions at $t=0$, one has $K-P=H_{\rm b}^{}+\epsilon=1.1$ ($K-P=\epsilon=0.1$). Thus, in the initial conditions of the second type, kinetic and potential energies of the system at $t=0$ are much closer than in the case of the first type of initial conditions.

It should be pointed out that the period of oscillations of the kinetic and potential energies is about $\pi/2\approx 1.57$ (see Fig.~\ref{fig:fig4}), while the period of temperature equilibration in the linear chain is much longer, since it is proportional to the chain length $N$; see Eq.~\eqref{tsimul}. For example, for a chain of $N=2^{10}=1024$ particles, the period of temperature equilibration is three orders of magnitude longer than the oscillation period of the kinetic and potential energies.

\begin{figure}[t!]
\includegraphics[width=8.0cm]{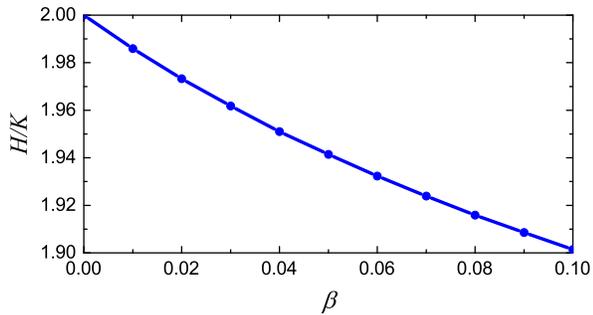}
\caption{Total to kinetic energy ratio as a function of the anharmonicity parameter $\beta=0$ for the case of total energy per particle equal to $H/N=1$, with the number of particles $N=2^{15}=32768$.}
\label{fig:fig4new}
\end{figure}

The main idea of this work is to analyze the effect of weak anharmonicity on the equilibration of sinusoidal temperature modulation, and thus it is important to quantify the proximity of the chain to the harmonic case. It is well known that in a harmonic chain, the kinetic energy is exactly half of the total energy, but this is not the case in the presence of anharmonicity. Let us use the total to kinetic energy ratio, $H/K$, as the measure of deviation from the harmonic case, which can also be used as the measure of heat capacity~\cite{Chetverikov1613,Chetverikov3815,Korznikova2020}. In Fig.~\ref{fig:fig4new}, the $H/K$ ratio is shown as a function of the anharmonicity parameter $\beta$ for the case of total energy per particle equal to $H/N=1$, with the number of particles $N=2^{15}$= 32768. The range $\beta\le 0.1$ is considered, which, as will be shown, is sufficient for our study; see Fig.~\ref{fig:btstar}. It follows from Fig.~\ref{fig:fig4new} that for $\beta<0.1$, the deviation of the $H/K$ ratio from the value corresponding to the harmonic limit ($\beta=0$) is within 5\%.

The presented analysis allows us to draw the following conclusions about the two types of the initial conditions. (i) The initial conditions of the second type are less stochastic (see Fig.~\ref{fig:s}), which means that they should give a better convergence of the results with increasing number of realizations. (ii) Simulations with longer chains should give a better convergence of the results with increasing number of realizations since $s$ decreases with increasing $N$; see Fig.~\ref{fig:s}. (iii) The initial conditions of the second type are closer to thermal equilibrium with closer values of the kinetic and potential energies in the system at $t=0$; see Fig.~\ref{fig:fig4}. (iv) For $\beta<0.1$, the total to kinetic energy ratio, $H/K$, differs by no more than 5\% from 2, which is the value for the harmonic case; see Fig.~\ref{fig:fig4new}. This means that for $\beta<0.1$, we have the regime of weak anharmonicity.

\subsection{Temperature equilibration in the linear chain}
\label{Linear}

\begin{figure}
\includegraphics[width=8.5cm]{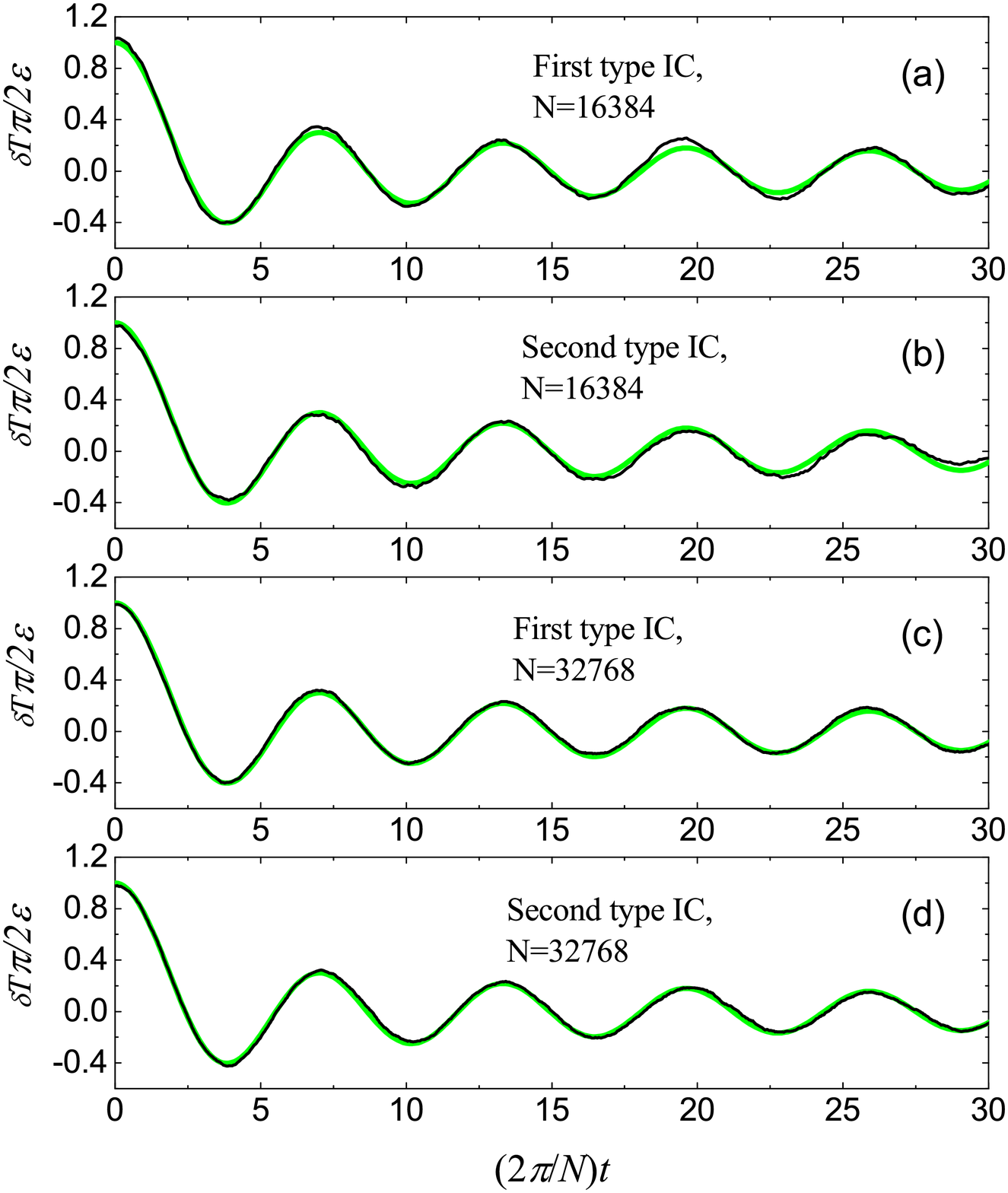}
\caption{Results for $\beta=0$ (linear case). The normalized difference between the averaged temperatures of the left and right halves of the chain, $\delta T$, is plotted as a function of normalized time. The numerical result is shown by a black line, while the theoretical prediction given by Eq.~\eqref{TLmTRfin} (the Bessel function of the first kind) is shown by the light green line. Initial conditions (a),(c) of the first type and (b),(d) of the second type are used. In (a),(b), the chain length is $N=16384$, and in (c),(d), it is $N=32768$. In all cases, the result of averaging over 50 realizations is shown.}
\label{fig:Linear}
\end{figure}

For the linear chain ($\beta=0$), the analytical result given by Eq.~\eqref{TLmTRfin} is available, according to which the temperature difference of the left and right halves of the chain oscillates with a reduction in the time amplitude following the Bessel function of the first kind. In Fig.~\ref{fig:Linear}, the normalized difference $\delta T=T_{\rm L}^{}-T_{\rm R}^{}$ as a function of the normalized time is plotted for $N=16384$ [Figs.~\ref{fig:Linear}(a) and \ref{fig:Linear}(b)] and $N=32768$ [Figs.~\ref{fig:Linear}(c) and \ref{fig:Linear}(d)]. In Figs.~\ref{fig:Linear}(a) and \ref{fig:Linear}(c), the initial conditions of the first type are used, while in Figs.~\ref{fig:Linear}(b) and \ref{fig:Linear}(d), those of the second type are used. Black curves show the numerical results and light green curves stand for the analytical prediction given by Eq.~\eqref{TLmTRfin}. In all cases, the result of averaging over 50 realizations is shown.

It can be seen from Fig.~\ref{fig:Linear} with the naked eye that the results for the longer chain shown in Figs.~\ref{fig:Linear}(c) and \ref{fig:Linear}(d) are closer to the theoretical prediction than the results for the shorter chain presented in Figs.~\ref{fig:Linear}(a) and \ref{fig:Linear}(b). In order to quantify the difference between the numerical and theoretical curves, we calculate the area $S$ between them within the interval $0\le(2\pi/N)t\le 30$. The result is (a) $S=0.7247$, (b) $S=0.7001$, (c) $S=0.4030$, and (d) $S=0.4027$. The smaller the area, the closer the numerical result is to the theoretical prediction. For (a) and (b), $S$ is considerably larger than for (c) and (d), and this is in line with the result presented in Fig.~\ref{fig:s}, showing that the standard deviation $s$ of $\delta T$ reduces with increasing chain length. It can also be seen that the area $S$ is slightly smaller when the initial conditions of the second type are used. This again agrees with Fig.~\ref{fig:s}, showing that $s$ is smaller for the second type of initial conditions.

From the results presented in Fig.~\ref{fig:Linear}, one can see that the number of realizations needed to achieve certain accuracy strongly depends on the chain length $N$. This is so because we monitor the time evolution of the integral parameter $\delta T$, and for a longer chain, this parameter is estimated with a higher accuracy for each particular realization. In the following, if the chain length is halved, the number of realizations is at least doubled to get approximately the same accuracy.

According to our simulations, for increasing number of realizations, the kinetics of temperature equilibration converges not only for the harmonic chain, but also for $\beta>0$.

It is clear that the theoretical result given by Eq.~\eqref{TLmTRfin} predicts the temperature equilibration in the linear chain very well. This confirms that the continuum equation \eqref{ContEq} is capable of describing the heat flux in linear chains.

\subsection{Effect of anharmonicity}
\label{Anharmonic}

\begin{figure}[t!]
\includegraphics[width=8.0cm]{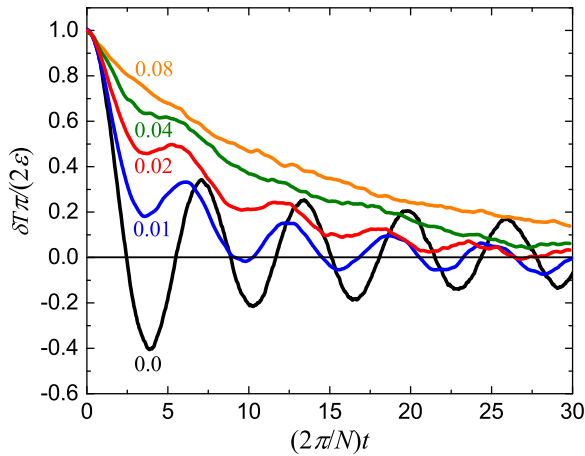}
\caption{Normalized $\delta T$ as a function of normalized time for different values of the nonlinearity parameter $\beta$, as specified for each curve. The initial conditions of the second type are used. The chain length is $N=32768$. All curves are the result of averaging over 50 realizations.}
\label{fig:NonLinear}
\end{figure}
\begin{figure}[t!]
\includegraphics[width=8.0cm]{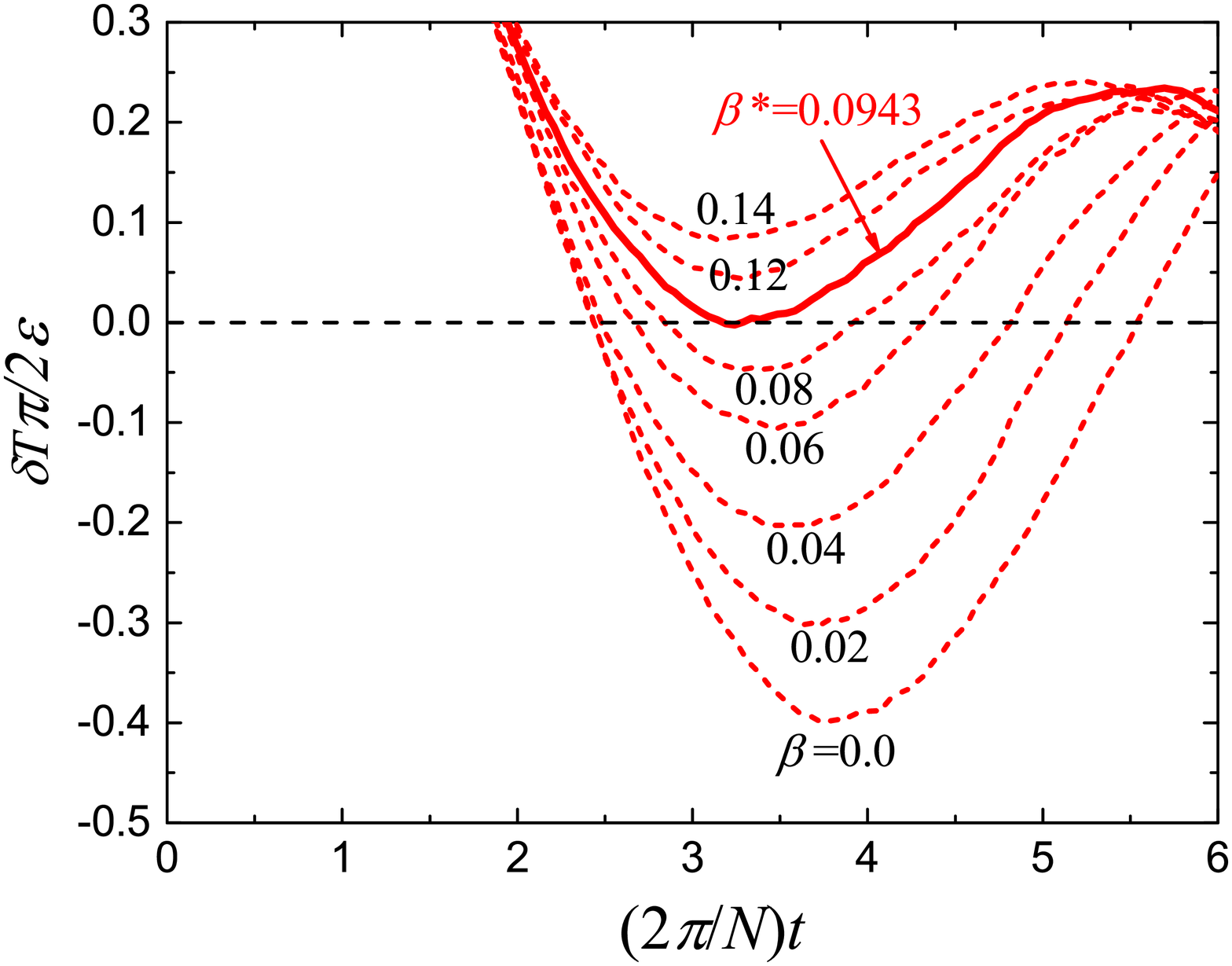}
\caption{A set of curves showing normalized $\delta T$ as a function of time for different values of the nonlinearity parameter $\beta$ in the vicinity of the first minimum. For $\beta^*=0.0943$, the first minimum of the $\delta T$ curve reaches zero value (shown by the solid line). The initial conditions of the first type are used. The chain length is $N=512$. All curves are the result of averaging over $2.5\times 10^4$ realizations. It is assumed that for $\beta<\beta^*$, the linear theory is still informative and, for larger $\beta$, the anharmonicity plays an essential role.}
\label{fig:star}
\end{figure}
\begin{figure}[t!]
\includegraphics[width=8.0cm]{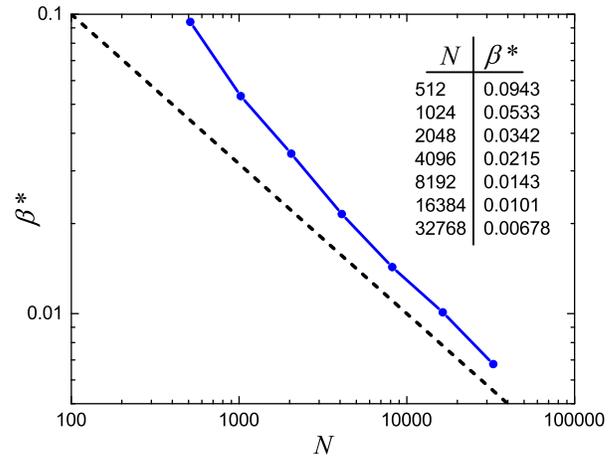}
\caption{The characteristic value of the anharmonicity parameter $\beta^*$ as a function of the wavelength of temperature modulation $N$. The dashed line shows the slope $-1/2$. The table shows the numerical values of $\beta^*$ for different $N$.}
\label{fig:btstar}
\end{figure}

The effect of the anharmonicity parameter $\beta$ on equilibration of the sinusoidal modulation of temperature is presented in Fig.~\ref{fig:NonLinear}. The time evolution of normalized $\delta T$ is presented and the value of $\beta$ is indicated for each curve. The chain length is $N=32768$. The results are averaged over 50 realizations. 

It can be seen from Fig.~\ref{fig:NonLinear} that with increasing $\beta$, the amplitude of oscillations of the curves decreases and, at $\beta=0.08$, one has an almost monotonically decreasing curve. The oscillation frequency increases with increasing $\beta$. It can also be noted that the fastest temperature equilibration is observed for $\beta=0.02$ because, in this case, the energy difference $\delta T$ vanishes faster. For smaller $\beta$, the amplitude of oscillations decays slower, and for larger $\beta$, the relaxation becomes slower, as compared to the case of $\beta=0.02$. We note the analogy with the damped oscillator, which relaxes to the equilibrium position very slowly if the viscosity is very high (overdamped motion) or when it is very small (slowly decaying oscillations), so that there exists a value of the viscosity parameter with the fastest relaxation~\cite{BookDamped}.

Our next step is to estimate the accuracy of the linear theory for different wavelengths of the temperature modulation, which, in our simulations, is equal to the chain lengths $N$. We need to define a characteristic value of the anharmonicity parameter $\beta^*$ such that for $\beta<\beta^*$, one can rely on the linear theory, but for large $\beta$, the effect of anharmonicity must be taken into account. 

As one can see from Fig.~\ref{fig:NonLinear}, with increasing $\beta$, the minima of the oscillating curve go up. Let us focus on the first minimum of the curve and specify $\beta^*$ as the value of $\beta$ when the first minimum of $\delta T$ reaches zero. This definition of $\beta^*$ is further illustrated in Fig.~\ref{fig:star}. Here a set of curves is plotted for different values of $\beta$ in the vicinity of the first minimum and interpolation between the minimal points allows one to find $\beta^*=0.0943$ when the first minimum of $\delta T$ vanishes (shown by the solid line). Note that for $\beta=\beta^*$, the oscillations of $\delta T$ in time are still pronounced and not yet suppressed by the anharmonicity. This means that for $\beta<\beta^*$, the linear theory is valid in the sense that it explains the oscillatory character of temperature equilibration in the chain. The results presented in Fig.~\ref{fig:star} are obtained for $N=512$, initial conditions of the first type are used, and averaging over $2.5\times 10^4$ realizations is performed for each value of $\beta$. Note that a larger number of realizations should be taken for short-wavelength temperature modulation.

Figure~\ref{fig:btstar} shows how the characteristic value of the anharmonicity parameter $\beta^*$ depends on the wavelength of the temperature modulation $N$. The dashed line in the log-log plot has the slope $-1/2$. This means that $\beta^*\sim N^{-1/2}$ for large $N$, and for smaller $N$ the decrease of $\beta^*$ with increasing $N$ is somewhat faster. The presented result says that the linear theory describes short-wavelength modulations of temperature in the chain with larger anharmonicity, while for long-wavelength modulations the effect of anharmonicity is stronger.

So far, the temperature modulation wavelength was equal to the chain length $N$. Let us check the effect of the chain length taking the modulation wavelength equal to 512 in the chain of 32768 particles, thus having 64 temperature modulation periods in the chain. In this case, the temperatures $T_{\rm L}^{}$ and $T_{\rm R}^{}$ are calculated as the sums over 64 half periods of temperature modulation. Simulations with the initial conditions of the first type have shown that for the long chain, $\beta^*=0.0963$, which should be compared to $\beta^*=0.0943$ found for the chain with a single period of temperature modulation (see Fig.~\ref{fig:btstar}). It is seen that the effect of the chain length is marginal because the characteristic value $\beta^*$ is nearly the same for the chains that include one and 64 periods of temperature modulation.

\section{Conclusions}
\label{Conclusion}

In the present study, the effect of anharmonicity on the equilibration of sinusoidal modulation of temperature in the $\beta$-FPUT chain was analyzed. The results for different values of the anharmonicity parameter $\beta$ and for different wavelengths of temperature modulation $N$ are obtained numerically and compared to the analytical solution available for the linear case ($\beta=0$). Applicability of the linear theory to a weakly nonlinear chain was assessed for different wavelengths of temperature modulation $N$. Initial conditions of two types were used: (i) at $t=0$, the energy of the system is in the form of kinetic energy with zero potential energy and (ii) the other, major part of the energy is initially shared between kinetic and potential energies.

Our main findings can be summarized as follows:
\begin{itemize}
	\item[(1)] For the linear chain ($\beta=0$), the numerical results averaged over increasing number of realizations converged to the analytical solution given by Eq.~\eqref{ContSolut}. This solution predicts that equilibration of sinusoidal modulation of temperature demonstrates oscillations with decrease in time amplitude, following the Bessel function of the first kind. This was true for the initial conditions of both types, though convergence with increasing number of realizations was faster for the initial conditions with nearly equal kinetic and potential energies. Convergence was also faster for larger wavelength of temperature modulation $N$; see Sec.~\ref{StatIniCond}. The kinetics of temperature equilibration, for increasing number of realizations, converges not only for the harmonic chain, but also for $\beta>0$.
	\item[(2)] With an increase in the degree of anharmonicity, the oscillatory equilibration of temperature gradually transforms into a monotonic one. For a given temperature wavelength modulation, there exists a value of the anharmonicity parameter when the temperature equilibration occurs most rapidly. For smaller values of $\beta$, oscillations of temperature decay slowly, and for larger $\beta$, the monotonic decay is slow; see Fig.~\ref{fig:NonLinear}. 
	\item[(3)] Linear theory remains informative for weakly anharmonic chains when $\beta<\beta^*$, with $\beta^*$ defined as shown in Fig.~\ref{fig:star}. As can be seen from Fig.~\ref{fig:btstar}, $\beta^*$ decreases with increasing temperature modulation wavelength $N$. This means that temperature modulation with short wavelength is less affected by the anharmonicity or, in other words, linear theory remains valid for larger values of $\beta$, as compared to the long-wavelength temperature modulation.
\end{itemize}

Overall, our results have confirmed that (i) the continuum equation \eqref{ContEq} derived in~\cite{Krivtsov407} accurately describes the temperature flow in linear chains, (ii) linear theory remains informative for weakly anharmonic chains, and (iii) short-wavelength modulations of temperature are less affected by the anharmonicity and linear theory remains valid for larger values of $\beta$, as compared to the long-wavelength modulations of temperature. 

In this regard, the results presented in previous works~\cite{Murad193109,Gendelman020103} have found their explanation. Oscillations of the short-wavelength sinusoidal temperature modulation, observed by the authors of those works, can be well explained by the linear theory~\cite{Krivtsov407}. The oscillations were not observed by the authors for long-wavelength temperature modulation because, in this case, the effect of anharmonicity is much stronger. The oscillations of long-wavelength temperature modulation can be observed for smaller values of the anharmonicity parameter.

\section*{Acknowledgments}

%\begin{acknowledgments}
The work of E.A.K.\ was supported by a grant of the President of the Russian Federation for State Support of Young Russian Scientists (Grant No.\ MD-3639.2019.2). The work of A.M.K.\ was financially supported by the Russian Science Foundation, Grant No. 18-11-00201. V.A.K.\ thanks the Russian Foundation for Basic Research for their financial support via Grant No.\ 20-37-70058. The work of D.X. was supported by NNSF (Grant No.\ 11575046) of China, and NSF (Grant No.\ 2017J06002) of Fujian Province of China. The work of V.A.G.\ was supported by the MEPhI Academic Excellence Project. S.V.D.\ acknowledges the support of the Russian Foundation for Basic Research, Grant No.\ 19-02-00971. 
%\end{acknowledgments}

\end{document}